# Exploring the complex interplay of anisotropies in magnetosomes of magnetotactic bacteria

*David Gandia[a], Lourdes Marcano[b,c] Lucía Gandarias[d,e], Alicia G. Gubieda[d], Ana García-Prieto[f], Luis Fernández Barquín[g], Jose Ignacio Espeso[g], Elizabeth Martín Jefremovas[g,h], Iñaki Orue[i], Ana Abad Diaz de Cerio[d], Mª Luisa Fdez-Gubieda[j] and Javier Alonso[g*]*

[a] Depto. Ciencias, Universidad Pública de Navarra, 31006 Pamplona, Spain

[b] Depto. de Física, Universidad de Oviedo, 33007 Oviedo, Spain

[c] CIC biomaGUNE, Basque Research and Technology Alliance (BRTA), Donostia-San Sebastián 20014, Spain

[d] Depto. Inmunología, Microbiología y Parasitología, Universidad del País Vasco (UPV/EHU), 48940 Leioa, Spain

[e] Aix-Marseille Institute of Biosciences and Biotechnologies (BIAM), Aix-Marseille Université, CNRS, CEA – UMR 7265, 13108 Saint-Paul-lez-Durance, France

[f] Depto. de Física Aplicada, Universidad del País Vasco (UPV/EHU), 48013 Bilbao, Spain

[g] Depto. CITIMAC, Universidad de Cantabria, 39005 Santander Spain

[h] Institute of Physics. Johannes Gutenberg University of Mainz, 55128 Mainz, Germany

[j] SGIker, Universidad del País Vasco (UPV/EHU), 48940 Leioa, Spain

[j] Depto. de Electricidad y Electrónica, Universidad del País Vasco (UPV/EHU), 48940 Leioa, Spain

*corresponding autor

E-mail addresses: david.gandia@unavarra.es (D. Gandia), marcanolourdes@uniovi.es (L. Marcano), lucia.gandarias@cea.fr (L. Gandarias), alicia.gascon@ehu.eus (A.G. Gubieda), ana.garciap@ehu.eus (A. García-Prieto), luis.fernandez@unican.es (L. Fernández Barquín), jose.espeso@unican.es (J. I. Espeso), martinel@uni-mainz.de (E.M. Jefremovas), inaki.orue@ehu.eus (I. Orue), ana.abad@ehu.eus (A. Abad Diaz de Cerio), malu.gubieda@ehu.eus (M. L. Fdez-Gubieda), javier.alonsomasa@unican.es (J. Alonso)

Magnetotactic bacteria (MTB) are of significant interest for biophysical applications, particularly in cancer treatment. The biomineralized magnetosomes produced by these bacteria are high-quality magnetic nanoparticles that form chains through a highly reproducible natural process. Specifically, *Magnetovibrio blakemorei* and *Magnetospirillum gryphiswaldense* exhibit distinct magnetosome morphologies: truncated hexa-octahedral and truncated octahedral shapes, respectively. Despite having identical compositions (magnetite, $Fe_3O_4$) and comparable dimensions, their effective uniaxial anisotropies differ significantly, with *M. blakemorei* showing ~25 kJ/m³ and *M. gryphiswaldense* ~11 kJ/m³ at 300K. This variation presents a unique opportunity to explore the role of different anisotropy contributions in the magnetic responses of magnetite-based nanoparticles.

This study systematically investigates these responses by examining static magnetization as a function of temperature (*M* vs. *T*, 5 mT) and magnetic field (*M* vs. $\mu_0 H$, up to 1 T). Above the Verwey transition temperature (~110 K), the effective anisotropy is dominated by shape anisotropy, notably increasing coercivity for *M. blakemorei* by up to two-fold compared to *M. gryphiswaldense*. Below this temperature, the effective uniaxial anisotropy increases non-monotonically, significantly altering magnetic behavior. Our simulations based on dynamic Stoner-Wohlfarth models indicate that below the Verwey temperature, a uniaxial magnetocrystalline contribution emerges, peaking at 22-24 kJ/m³ at 5 K—values close to those of bulk magnetite. This demonstrates the profound impact of anisotropic properties on the magnetic behaviors and applications of magnetite-based nanoparticles and highlights the exceptional utility of magnetosomes as ideal model systems for studying the complex interplay of anisotropies in magnetite-based nanoparticles.

## 1. Introduction

Magnetotactic bacteria (MTB) are unique aquatic microorganisms capable of directional movement guided by Earth's magnetic field [1–3]. This remarkable ability is due to the presence of intracellular magnetosomes. The latter are magnetic nanoparticles enclosed in a lipid bilayer, which tend to form chains inside the MTB, and are predominantly composed of magnetite ($Fe_3O_4$). Interestingly, these nanoparticles demonstrate a high degree of chemical purity and uniformity in size (35-120 nm), achieved through a naturally-driven biomineralization process that ensures reproducibility and magnetic stability of the magnetosomes at room temperature. [4,5].

Recent years have witnessed burgeoning interest in MTB, mainly driven by the potential of magnetosomes in various biomedical applications, such as Magnetic Hyperthermia and Magnetic Resonance Imaging (MRI), and by the possibility of using MTB as remotely controlled microrobots [6–21]. Beyond their practical applications, magnetosomes serve as excellent subjects for fundamental nanomagnetic studies due to their consistent composition, size, shape, and arrangement. This precision allows them to model nanomagnetic phenomena, like the influence of dipolar interactions in one-dimensional assemblies or the effect of particle morphology on magnetic response at the nanoscale [22–25]. To this respect, our previous research utilized experimental techniques and computational methods to highlight the significant role of particle shape in determining the magnetic properties of faceted magnetosomes, specifically those synthesized by *M. gryphiswaldense* [25]. These studies revealed that magnetosomes from this species presented a slight deformation which tilted the magnetic moment out of the chain easy axis and allowed to accommodate the chain to the characteristic helical morphology of these bacteria. This deformation in the truncated octahedral shape of these magnetosomes crucially influence their magnetic response and heating efficiency in therapeutic applications like Magnetic Hyperthermia [26].

However, leveraging magnetosomes as model systems presents challenges, primarily because their morphology, governed by genetic factors, is species-specific and not readily modifiable like synthetic nanoparticles [27–31]. The diversity of MTB species and their magnetosome morphologies is vast, yet culturing these organisms in a laboratory setting remains a significant hurdle [32–35], limiting most research to a few species like *Magnetospirillum gryphiswaldense* MSR-1 and *Magnetospirillum magneticum* AMB-1 [36,37].

This study goes a step beyond by culturing and determining the anisotropy variations in *Magnetovibrio blakemorei* strain MV-1, which produces elongated $Fe_3O_4$ magnetosomes with a distinct truncated hexa-octahedral shape. This allows for a direct comparison of the magnetic properties with *M. gryphiswaldense*. First isolated by Bazylinski *et al.* in 1988 [38], *M. blakemorei* has been less studied due to the challenges associated with its cultivation [2,39–43].

This work presents a detailed analysis of the magnetic properties of *M. blakemorei*, comparing them with those of *M. gryphiswaldense*. Both species synthesize high-quality $Fe_3O_4$ magnetosomes with comparable size but different shape. Our experiments, conducted with the magnetosomes intact within the bacteria, allow precise control over magnetic interactions and external conditions, enabling us to simulate and compare their magnetic responses under various temperatures. Utilizing experimental measurements focused on the collection of hysteresis loops across a wide range of temperatures (5 to 300 K), along with calculations based on a dynamic Stoner-Wohlfarth model, the thermal evolution of their magnetic anisotropies has been dissected, providing invaluable insights into the fundamental magnetic parameters that govern the behavior of magnetite-based nanoparticles of biotechnological interest. This study not only advances our understanding of MTB but also underscores the significance of magnetosomes as ideal model systems in nanomagnetic research, potentially benefiting various applications in the field of magnetic nanoparticles, not only restricted to biomedicine, but also expanding to emergent fields, like topological spin textures or bio-encoded magnonics [44].

## 2. Experimental

*2.1 Magnetotactic bacteria culture*

*Magnetospirillum gryphiswaldense* MSR-1 (DMSZ 6631) was grown without shaking at 28 °C in flask standard medium using 0.3% (wt/vol) of sodium pyruvate as carbon source and supplemented with 100 µM of Fe(III)-citrate (REF). [45].

*M. blakemorei* strain MV-1 (DSM 18854) was grown anaerobically at 30°C in liquid medium containing per liter of artificial seawater (ASW): 4 g Na succinate x $6H_2O$, 0.8 g Na acetate, 1 g casamino acids (BD Bacto), 1 g $NH_4Cl$, 5 mL Wolfe's mineral solution and 50 μL of 1%(w/v) resazurine. After autoclaving, 0.5 mL BME Vitamins 100x solution (Sigma-Aldrich, B6891), 1.8 mL 0.5 M $PO_4$ buffer, 0.3 mL Fe(II)-$Cl_2$ and 10 mL of 0.25 M freshly made cysteine solution were added to the media and the pH was adjusted to 7. The media was then distributed into sterile Hungate tubes and fluxed for 20 minutes with $N_2O$.

Once bacteria present well-formed magnetosome chains, the cells were harvested by centrifugation, fixed with 2% glutaraldehyde, washed three times with PBS and finally suspended in Milli-Q water.

### *2.2 Transmission Electron Microscopy*

Transmission electron microscopy (TEM) was carried out on whole bacteria (i.e., whole cells) deposited onto 300 mesh carbon-coated copper grids. The images were obtained with a JEOL JEM-14000 Plus electron microscope at an accelerating voltage of 120 kV. The particle size distribution was analyzed with ImageJ software [46].

### *2.3 Magnetic measurements*

The magnetic characterization was carried out using the whole bacteria. The samples were freeze-dried and encapsulated in gelatin capsules. Magnetic measurements were performed in a superconducting quantum interference device magnetometer (Quantum Design MPMS-5). Magnetization vs. temperature (*M* vs. *T*) curves were measured following the usual zero-field-cooling/field-cooling (ZFC/FC) protocol, with an applied magnetic field of 5 mT. Magnetization vs. magnetic field (*M* vs. µ₀*H*) loops were measured both in ZFC and FC modes, at different temperatures, 5–300 K, applying magnetic fields, µ₀*H*, up to 1 T.

## 3. Results and Discussion

**Figure 1** shows representative TEM images of the *M. blakemorei* and *M. gryphiswaldense* bacteria, and their chain of magnetosomes. As depicted, both species synthesize a single chain of magnetosomes, but in the case of *M. blakemorei,* these magnetosomes present an elongated truncated hexa-octahedral shape (see the inset to **Figure 1**) which differs from the truncated octahedral shape typically obtained for *M. gryphiswaldense* and other species of the genus *Magnetospirillum*. In both cases, the magnetosomes inside the chain are aligned along a [111] crystallographic direction of magnetite, which defines the so-called *chain axis* [22]. Concerning the chain of magnetosomes, it is noticeable that the length of the chain is larger in the case of *M. gryphiswaldense* with an average of 15(3) magnetosomes per chain, while in the case of *M. blakemorei*, the average number of magnetosomes reduces to 9(2). On the other hand, the dimensions of the magnetosomes from *M. blakemorei*, as determined from the corresponding TEM images, are 45(4) × 45(4) × 65(5) nm³, while the truncated octahedral magnetosomes from *M. gryphiswaldense* have an average size of 45(5) nm. Despite the differences in their morphology, both magnetosomes present comparable dimensions.

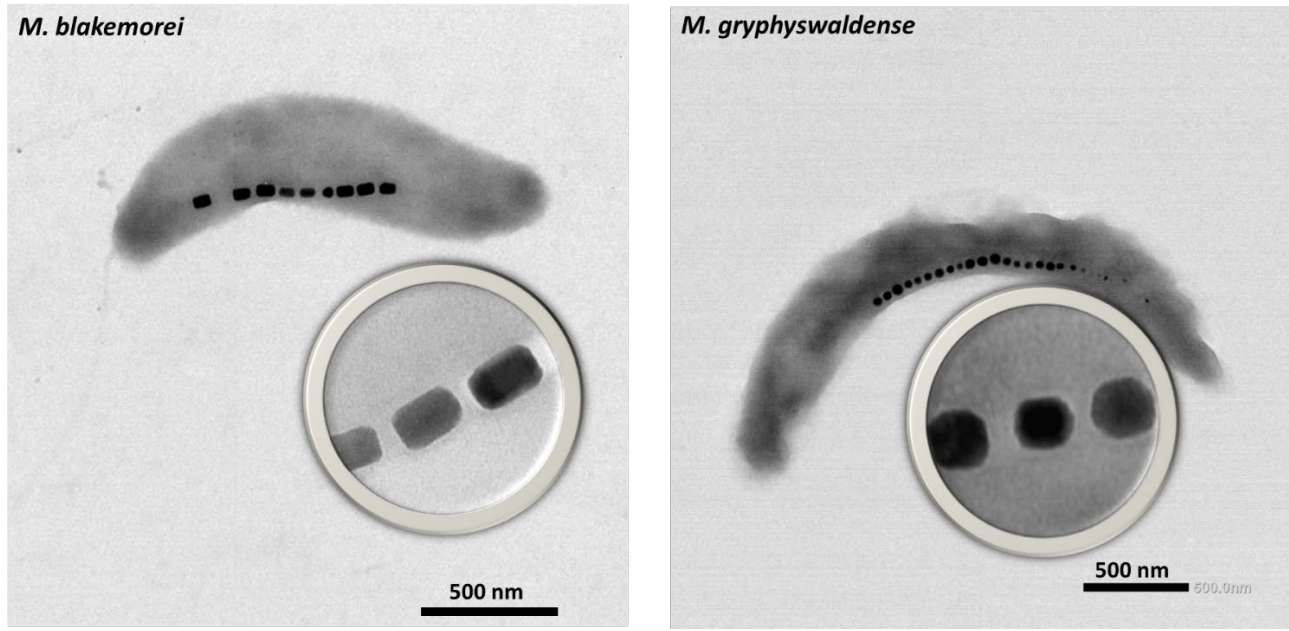

**Figure 1.** TEM images of the *M. blakemorei* and *M. gryphiswaldense* bacteria. The chain of magnetosomes can be clearly identified inside the bacteria. In the insets, a zoom-in of the chain is presented to showcase the different morphology of the magnetosomes from the two species.

The different morphology of both strains results, naturally, in different shape anisotropy contributions. The shape anisotropy energy landscape has been calculated using a Finite Element Method (FEM) protocol described in detail before (see Refs [23,25]), and the main results are summarized in **Figure 2.** This figure presents a schematic depiction of the morphology of each magnetosome, with the corresponding crystallographic axes and shape anisotropy energy landscape. In Ref [25], by using FEM, we showed that the magnetosomes synthesized by *M. gryphiswaldense* present a slight deformation that gives rise to a deformed toroidal energy landscape, with a unique quasi-uniaxial character (i.e., a main uniaxial contribution + a smaller cubic contribution). This quasi-uniaxial easy axis is effectively tilted ~20° out of the [111] chain axis, as represented in **Figure 2**, and the corresponding shape anisotropy constant values are $K_{sh-uni} \sim 7$ kJ m$^{-3}$ and $K_{sh-cub} \sim 1.5$ kJ m$^{-3}$ at 300 K. Instead, in the case of *M. blakemorei* [23], the shape anisotropy energy landscape exhibits a perfect toroidal shape indicating the presence of a well-defined uniaxial easy axis along the chain axis direction. These magnetosomes exhibit some dispersion in their aspect ratio, i.e., width/length (W/L), with an average value: W/L = 0.70(6), which gives rise to a uniaxial shape anisotropy constant value of $K_{sh-uni} \sim 23(4)$ kJ m$^{-3}$ at 300 K. Therefore, these results clearly display that the elongated morphology of *M. blakemorei* magnetosomes gives rise to a stronger and strictly uniaxial shape anisotropy compared to *M. gryphiswaldense*. This is going to have an important impact on their magnetic response, as we will see below.

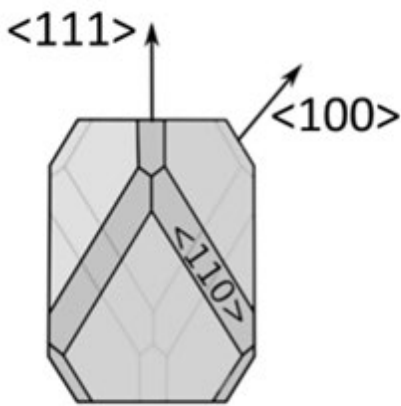
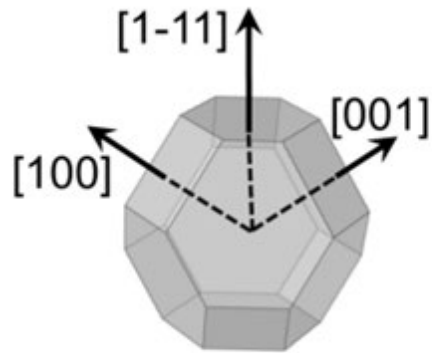
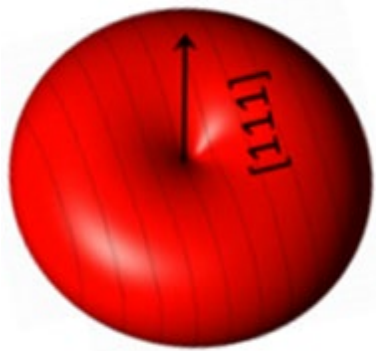
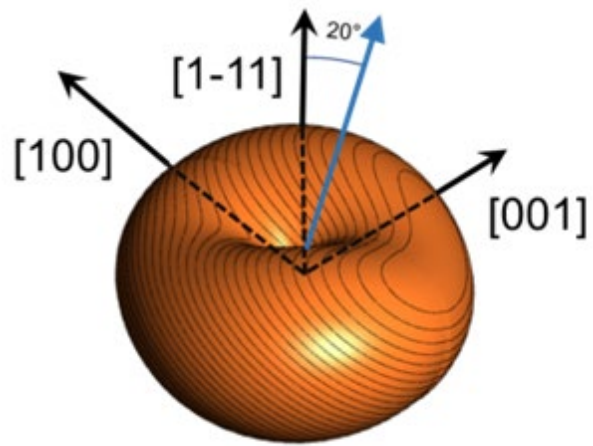

**Figure 2.** (Top) Schematic depiction of the morphology of the magnetosomes of *M. blakemorei* (left) and *M. gryphiswaldense* (right), including the crystallographic axes. The [111] direction corresponds with the chain axis of the MTB. (Bottom) Corresponding shape anisotropy energy landscapes, which have been calculated following the procedure described in Gandia *et al* [25].

To probe the influence of the different anisotropy energy landscapes on the magnetic properties, static magnetic measurements have been carried out, both as a function of the temperature (*M* vs. *T*) and as a function of the magnetic field (*M* vs. $\mu_0 H$), from 5 to 300 K. **Figure 3 a)** showcases the *M* vs. *T* curves measured in the ZFC/FC mode at low fields, 5 mT, for both species. These ZFC/FC curves present a qualitatively similar behavior in both cases: there is a strong irreversibility between the ZFC and FC curves, and a sharp transition appears in both curves around 100-110 K, which corresponds to the well-known Verwey transition of magnetite [47–49]. As has been explained before, this is a temperature phase transition in which magnetite crystal lattice changes from cubic to monoclinic with decreasing temperature, and the magnetocrystalline anisotropy changes from cubic to uniaxial. The presence of the Verwey transition in the ZFC/FC curves is a clear indicator of the magnetosomes being made of $Fe_3O_4$. In bulk magnetite, this transition gives rise to a sharp step around $T_v = 120$ K [50,51]. However, in magnetosomes, and in other high quality magnetite nanoparticles, this transition tends to be slightly shifted and smoothed. This can be in principle related to the stoichiometry moving away from that of pure magnetite due to the presence of vacancies, defects, or doping elements [52–54]. In our case, the Verwey temperature is shifted ~ 10 K towards lower temperatures for both species. This can be more easily discerned if we inspect the first derivative of the ZFC curves (**Figure 3 b)**), where a well-defined peak appears around 90-100 K, corresponding to the Verwey transition. This peak is slightly sharper for *M. blakemorei*, suggesting a more homogenous stoichiometry. This is also supported by the value of the onset of the Verwey temperature (marked with black arrows): $T_v \sim 112$ K for

*M. blakemorei*, slightly closer to the expected bulk value, $T_v \sim 120$ K, than for *M. gryphiswaldense*, $T_v \sim 107$ K. In addition, as marked by the change in the slope of the ZFC curves and the second peak in the derivative, a second transition appears around 30-50 K. These correspond to the so-called "low-*T* transitions", frequently reported in bulk magnetite and magnetite-based nanoparticles, although their nature is still under debate [55–60]. As depicted in the derivative, there are some small differences in the shape and position (marked by white arrows) of this low *T* peak, around $T_{low-T} \sim 27$ K and $\sim 30$ K for *M. blakemorei* and *M. gryphiswaldense*, respectively. Therefore, these results indicate similar crystallographic transitions in both magnetosomes despite their marked difference in morphology.

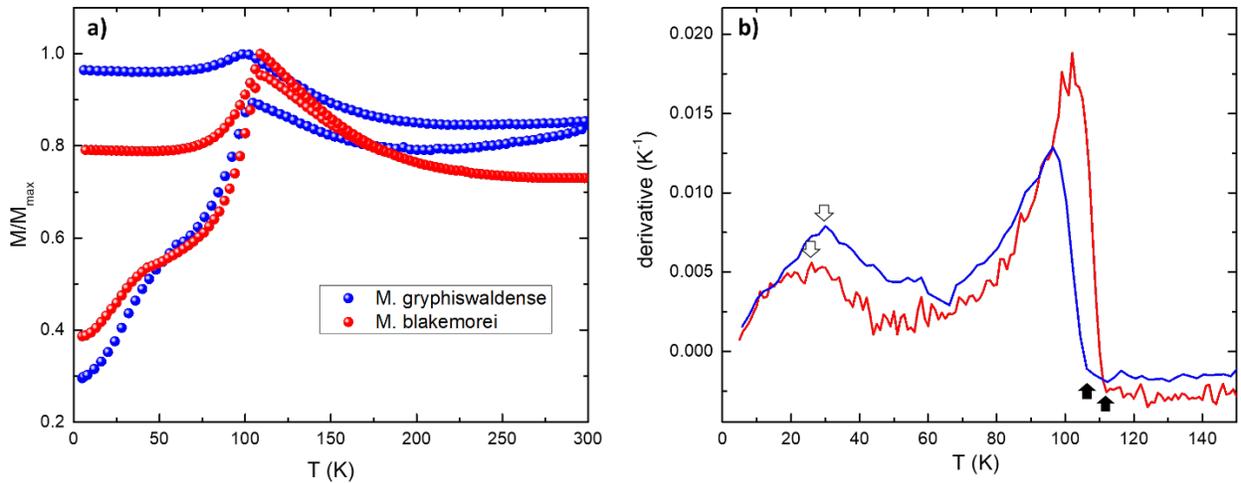

**Figure 3.** a) ZFC/FC curves measured at 50 Oe for *M. blakemorei* and *M. gryphiswaldense*. b) First derivative of the ZFC curve (5-150 K). Arrows in the plot mark the two transitions discussed in text, i.e. low-*T* transition (white arrows) and Verwey transition (black arrows).

Next, in order to gain a better insight into the complex interplay of anisotropies in these two MTB species, their magnetic response as a function of the applied magnetic field has been analyzed by measuring the *M* vs. $\mu_0 H$ loops up to 1 T at different temperatures between 5 and 300 K, both in ZFC and FC modes for *M. blakemorei* (see **Figure S1** in Supporting Information). Briefly, above

$T_v$, the ZFC and FC hysteresis loops are nearly identical, but as the temperature decreases, the FC loops show lower coercivity and higher remanence than the ZFC loops; additionally, *M. blakemorei* exhibits wider loops with higher irreversibility than *M. gryphiswaldense*, indicative of higher effective anisotropy. To better analyze the differences, in **Figure 4** we have represented the thermal evolution of the coercive field, $\mu_0 H_c$ (left), and the normalized magnetization remanence, $M_r/M_s$ (right) for both species.

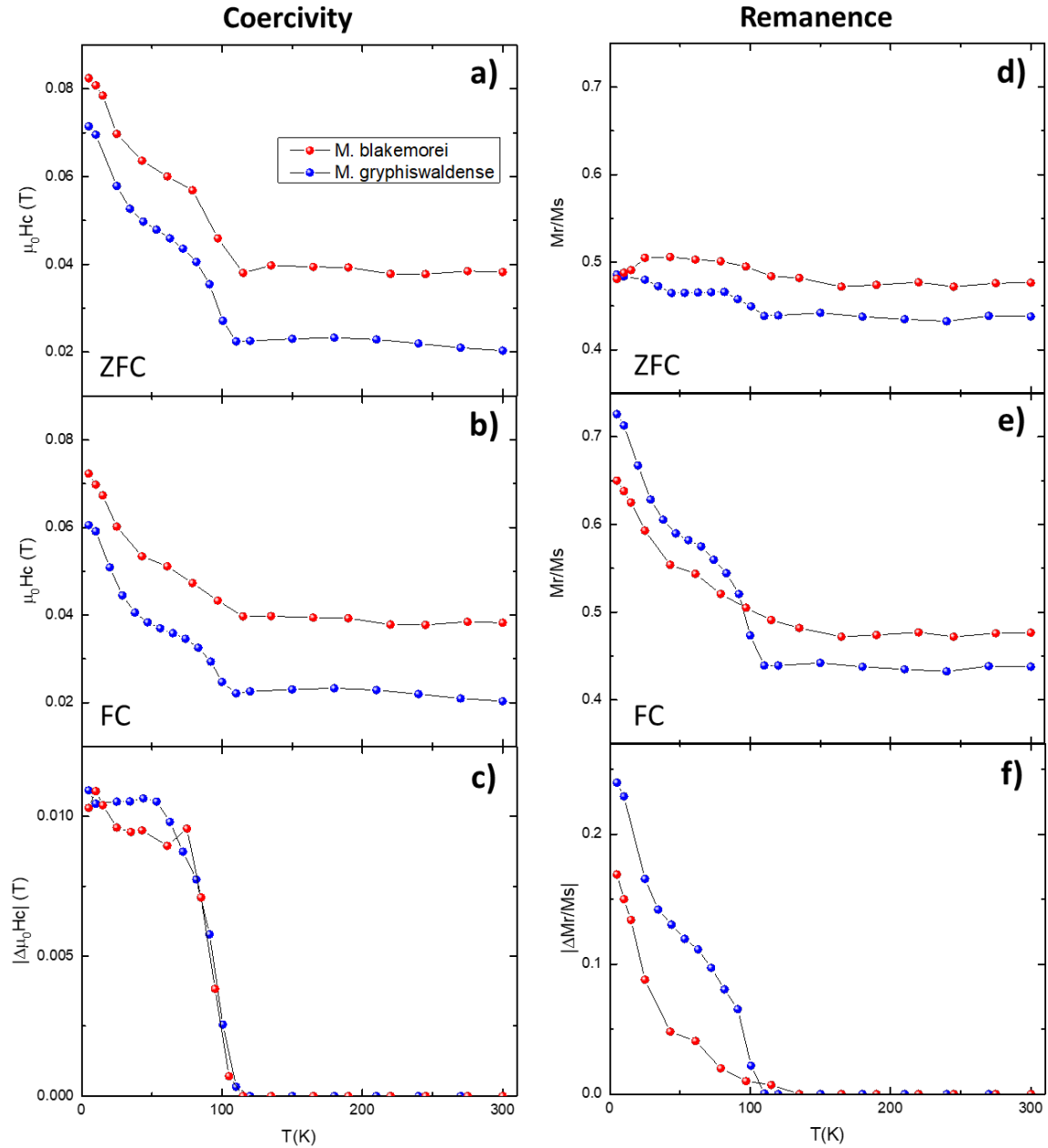

**Figure 4.** Thermal evolution of the a) ZFC and b) FC Coercive field, $\mu_0H_c$, d) ZFC and e) FC normalized remanence, $M_r/M_s$, vs temperature curves obtained from the corresponding $M$ vs. $\mu_0H$ loops. The difference, in absolute value, between these FC and ZFC curves is represented in c) for the coercive field, $|\Delta \mu_0H_c|$, and in f) for the normalized remanence, $|\Delta M_r/M_s|$.

The evolution of $\mu_0H_c$ with temperature (**Figures 4 a), b)**) is qualitatively similar for both bacteria species, but the obtained values are appreciably higher (up to 2 times greater) for *M. blakemorei*.

This enhanced coercivity can be directly associated to the higher shape anisotropy of *M. blakemorei*, since it is proportional to the anisotropy: $H_c = 2K/\mu_0 M_s$ for single domain uniaxial nanoparticles, following a *coherent rotation* or *Stoner-Wohlfarth* mode during the reversal of the magnetization [61]. This shows that modifying the shape of MNPs is a powerful strategy to tune the coercive field, which is especially interesting in different applications such as magnetic hyperthermia or data storage [28,30,62,63]. Concerning the thermal evolution, we can see that $\mu_0 H_{c\text{-ZFC}}$ remains nearly constant ($\mu_0 H_{c\text{-ZFC}}$ = 40 mT and 20 mT for *M. blakemorei* and *M. gryphiswaldense*, respectively) from 300 K down to ~110 K, that is, $T_v$. In the case of bulk magnetite, the coercive field steeply increases below $T_v$, and then remains nearly constant [64], which is related to the sharp transition taking place: with decreasing temperature, the magnetocrystalline anisotropy changes from cubic ($K_{cub}$ = -11 kJ/m$^3$ and ⟨111⟩ easy axes) to essentially uniaxial ($K_{uni}$ = +25 kJ/m$^3$ and ⟨100⟩ easy axes) as the crystalline structure evolves from cubic to monoclinic, giving rise to a sudden increase in the coercive field [65]. However, in the case of magnetosomes and many other magnetite-based nanoparticles [52,66,67], this process is, unsurprisingly, less abrupt and more complex than in bulk. Here, we can see $\mu_0 H_{c\text{-ZFC}}$ first increases below $T_v$, then there is a shoulder around 70-75 K, and finally increases again below 40-45 K, reaching a maximum value at 5 K of 80 mT and 70 mT for *M. blakemorei* and *M. gryphiswaldense*, respectively. A similar behavior, albeit smoother, is obtained for the $\mu_0 H_{c\text{-FC}}$ curves (**Figure 4 b)**). The differences between both MTB can be better tracked by representing the relative change of the coercive field, $|\Delta\mu_0 H_c|$ vs $T$ curves (**Figure 4 c)**). As depicted, the curves nearly overlap for both samples, indicating that, despite the difference in coercive field values, the magnetic response and the magnetic transitions taking place are very similar. However, there are minor differences below 70 K, which indicate some changes in the magnetocrystalline anisotropy of both samples, as had already been showcased in the $M$ vs. $T$ curves.

Focusing now on the normalized remanence, on the one hand, it can be observed that the ZFC $M_r/M_{S\text{-ZFC}}$ (**Figure 4 d)**) remains fairly constant for both MTB species: for *M. gryphiswaldense*, $M_r/M_{S\text{-ZFC}} \sim 0.44$ from 300 K down to 110 K, and then it increases up to $M_r/M_{S\text{-ZFC}} \sim 0.49$ at 5 K; while for *M. blakemorei*, $M_r/M_{S\text{-ZFC}} \sim 0.48$ from 300 K down to 110 K, and then increases up to 0.50. These values are close to ~0.5, indicating that, overall, these magnetosomes behave like randomly-oriented uniaxial single domain magnetic nanoparticles, as typically described in the framework of the Stoner-Wohlfarth model [68]. This behavior is more closely followed by the elongated magnetosomes from *M. blakemorei*, as expected, due to their better-defined uniaxial shape anisotropy. On the other hand, in the case of the FC remanence values, $M_r/M_{S\text{-FC}}$, (**Figure 4 e)**) we can see, however, some clear changes. For *M. gryphiswaldense*, $M_r/M_{S\text{-FC}}$ increases abruptly from 0.44 below the Verwey transition up to 0.72 at 5 K. Noticeably, a shoulder around 60 K, resembling the one observed in the $\mu_0 H_c$ vs $T$ curves, is also present in these $M_r/M_{S\text{-FC}}$ vs $T$ curves. On the other hand, for *M. blakemorei*, we see a more progressive evolution: $M_r/M_{S\text{-FC}}$ remains nearly constant, ~0.48, from 300 K to 160 K, and then it increases up to 0.65 at 5 K. The relative change of the normalized remanence, $|\Delta M_r/M_S|$ vs. $T$ (**Figure 4 f)**) allows to better detect the differences between both MTB. Above $T_v$, $|\Delta M_r/M_S|$ is null for both species, but below $T_v$, a progressive increase in $|\Delta M_r/M_S|$ is obtained, more abrupt in the case of *M. gryphiswaldense*.

Therefore, these results indicate that, despite the similarities between both MTB, there are some clear differences in their magnetic response, which can be related to the various anisotropies influencing the magnetic behavior of these bacteria. To fully capture the physics underneath, computing simulations of the experimental $M$ vs. $\mu_0 H$ loops have been carried out. These simulations are based on a magnetic model that has been described before in detail in previous studies [69,70]. Each magnetosome in the chain is considered as a single domain magnetic moment. For a given magnetic field at a fixed temperature, the equilibrium configuration of these magnetic moments is determined calculated by minimizing the single dipole energy density, given as the sum of three contributions: (i) the effective cubic anisotropy energy density ($E_{cub}$), due to the magnetocrystalline anisotropy above $T_v$ and the small cubic contribution to the shape anisotropy for *M. gryphiswaldense*; (ii) the effective uniaxial anisotropy energy density ($E_{uni}$), arising from the competition between magnetosome shape anisotropy and dipolar interactions within the chain (across all temperatures), including the magnetocrystalline anisotropy contribution below $T_v$; and (iii) the Zeeman energy density term ($E_Z$). Considering, $\theta$ and $\varphi$ the polar and azimuthal angles of the magnetic moment of each magnetosome, with the $\langle 100 \rangle$ magnetite crystallographic directions as reference, the total energy density is given by:

$$E(\theta, \varphi) = E_{cub}(\theta, \varphi) + E_{uni}(\theta, \varphi) + E_Z(\theta, \varphi) \qquad (1)$$

where:

$$E_{cub}(\theta, \varphi) = \frac{K_{cub}}{4}[\sin^4(\theta)\sin^2(2\varphi) + \sin^2(2\theta)] \qquad (2)$$

$$E_{uni}(\theta, \varphi) = K_{uni}[1 - (\hat{u}_m \cdot \hat{u}_{uni})^2] \qquad (3)$$

$$E_Z(\theta, \varphi) = -\mu_0 M_S H (\hat{u}_m \cdot \hat{u}_H) \qquad (4)$$

being $K_{cub}$ the effective cubic anisotropy constant; $K_{uni}$ the effective uniaxial anisotropy constant; $\hat{u}_m$ the magnetic moment unit vector; $\hat{u}_{uni}$ the unit vector along the effective uniaxial easy axis; $\hat{u}_H$ the external magnetic field unit vector; and $M_s$ the spontaneous magnetization of magnetite,

$M_s = 48 \times 10^4$ A/m. As we mentioned before (see **Figure 2**), the magnetosomes from *M. gryphiswaldense* present a quasi-uniaxial easy axis, $\hat{u}_{uni}$, which is tilted ~20° out of the [111] chain axis, due to a slight deformation of their truncated octahedral shape [25]. However, for the magnetosomes from *M. blakemorei*, $\hat{u}_{uni}$ is clearly pointing along the [111] chain axis due to their elongated shape. Taking into account these considerations, the *M* vs. $\mu_0 H$ loops at different angles have been simulated, using a dynamic approach described by Carrey *et al* [70], and averaged in order to reproduce the experimentally obtained ZFC *M* vs. $\mu_0 H$ loops corresponding to randomly oriented MTB. In this way, for each loop at a certain temperature, $K_{cub}$ and $K_{uni}$ have been adjusted to attain the best match between experimental and simulated *M* vs. $\mu_0 H$ loops.

The experimental and simulated ZFC *M* vs. $\mu_0 H$ loops obtained at 5, 115, and 300 K for *M. blakemorei* and *M. gryphiswaldense* using this model have been represented in **Figure 5 a)**. As depicted, an excellent match between experimental values and simulations has been reached in all cases by properly tuning the values of $K_{cub}$ and $K_{uni}$. Similar results have been obtained for the rest of the temperatures. These have allowed us to plot the thermal evolution of both anisotropy constants, as depicted in **Figures 5 b), c)**. Interestingly, while in the case of *M. gryphiswaldense* both $K_{cub}$ and $K_{uni}$ need to be included in the model in order to obtain a good fitting above $T_v$, in the case of *M. blakemorei*, only $K_{uni}$ needs to be taken into consideration in the whole temperature range. For the later, the strong uniaxial anisotropy energy, $E_{uni}$, in the [111] direction makes the cubic magnetocrystalline anisotropy energy, $E_{cub}$, irrelevant for these calculations, and the shape of the calculated *M* vs. $\mu_0 H$ loops does not change by including the $E_{cub}$ term. Therefore, we have not considered this term for *M. blakemorei*, and only the effective uniaxial anisotropy term, $E_{uni}$, has been included in the simulations of this species, with a Gaussian distribution to take into account the aspect ratio dispersion existing in these specific magnetosomes (i.e., W/L = 0.70(6)). This explains why $K_{cub}$ is only represented for *M. gryphiswaldense* in **Figure 5 b)**.

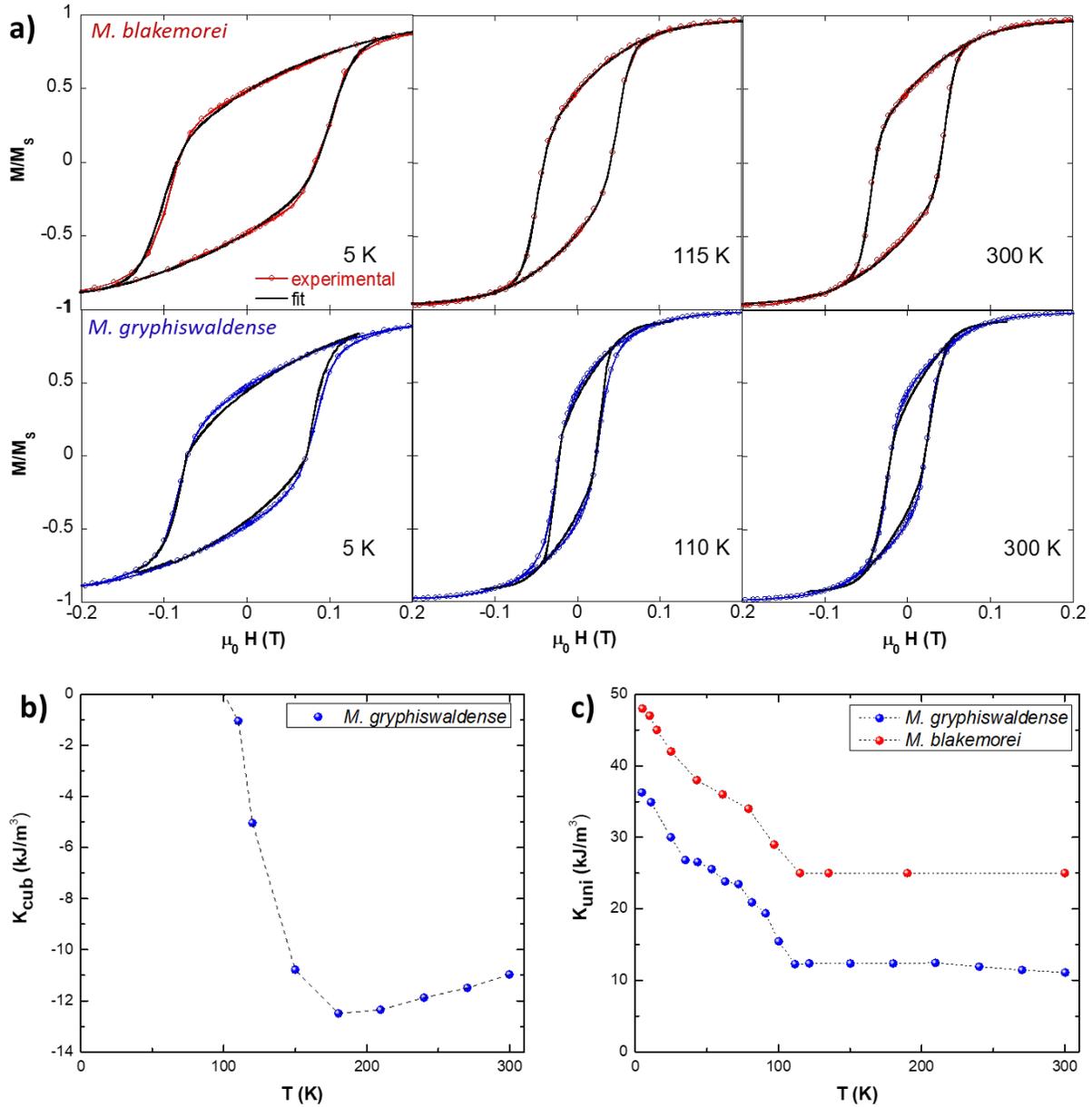

**Figure 5.** a) Experimental and simulated ZFC $M$ vs. $\mu_0H$ loops at 5, 115, and 300 K obtained for *M. blakemorei* and *M. gryphiswaldense*. Thermal evolution of the effective b) cubic anisotropy ($K_{cub}$) and c) uniaxial anisotropy ($K_{uni}$) as obtained from the simulations.

The thermal evolution of the cubic contribution to the anisotropy, $K_{cub}$, is mainly related to the magnetocrystalline anisotropy. Although it is true that magnetosomes from *M. gryphiswaldense* present a non-negligible cubic contribution to the shape anisotropy, $K_{sh-cub}$ = 1.5 kJ m$^{-3}$ at 300 K [25], this value is nearly one order of magnitude smaller than the magnetocrystalline contribution. In **Figure 5 b)**, it can be observed that at room temperature, the obtained value of $K_{cub}$ is around -11.0 kJ m$^{-3}$, close to the expected $K_{cub}$ value for bulk magnetite, -10.8 kJ m$^{-3}$. This serves as a reminder of the well-known high structural quality of these magnetosomes. With decreasing temperature, $K_{cub}$ slowly increases in absolute value up to -12.5 kJ m$^{-3}$ at 180 K. After that, it rapidly decreases in absolute value until becoming null at 110 K, around the Verwey transition.

Concerning the thermal evolution of the uniaxial contribution, $K_{uni}$, it is important to recall that, above the Verwey temperature, it gathers two contributions: the shape uniaxial anisotropy ($K_{sh-uni}$) and the dipolar interactions between magnetosomes inside the chain ($K_{dip}$). On the other hand, below the Verwey temperature, a third contribution should appear, associated with magnetocrystalline anisotropy, as explained before. In **Figure 5 b)**, already at 300 K, we can see some clear differences between the two species analyzed in this work: the $K_{uni}$ value for *M. blakemorei* (~ 25 kJ m$^{-3}$) is appreciably higher than the one obtained for *M. gryphiswaldense* (~ 11 kJ m$^{-3}$). Since the values for the shape anisotropy of both magnetosomes at 300 K have been already calculated using our FEM model [25] (~7 kJ m$^{-3}$ for *M. gryphiswaldense* and ~23 kJ m$^{-3}$ for *M. blakemorei*), we can directly obtain an estimation of the effective dipolar anisotropy ($K_{dip}$ = $K_{uni}$ - $K_{sh-uni}$), being $K_{dip}$ ~4 kJ m$^{-3}$ for *M. gryphiswaldense* and ~2 kJ m$^{-3}$ for *M. blakemorei*.

From 300 K to 110 K, i.e., down to the Verwey transition, $K_{uni}$ remains almost constant for both species (there is a slight increase in the case of *M. gryphiswaldense*). This indicates that the thermal dependence of both shape and dipolar anisotropies, in this range of temperatures, is relatively small. However, below the Verwey temperature, $K_{uni}$ increases from 12 to 36 kJ m$^{-3}$ at 5 K for *M. gryphiswaldense*, and from 25 to 48 kJ m$^{-3}$ at 5 K in the case of *M. blakemorei*. The evolution of $K_{uni}$ vs $T$ below the Verwey transition follows the same qualitative trend we saw for $\mu_0 H_c$ vs $T$: a first increase with a shoulder between 75-35 K, approximately, and a second increase below that temperature down to 5 K. This increase in $K_{uni}$ is associated, as explained before, with the change in magnetocrystalline anisotropy, from cubic to uniaxial.

Concerning the shape and dipolar anisotropies, as it is well known, their thermal dependence should be proportional to thermal evolution of $M_S^2$ [61]. Our results indicate that above the Verwey temperature, the saturation magnetization roughly follows a Bloch-like law, $M_s(T) \propto M_0[1 - \left(\frac{T}{T_0}\right)^{\alpha_B}]$, being $M_0$ the saturation magnetization at 0 K, $T_0$ the temperature at which $M_s$ becomes null, i.e. the Curie temperature (~843 K for magnetite), and $\alpha_B$ is a Bloch-like exponent (see **Figure S2**). In our case $\alpha_B$ acquires a value ~2.65. Note that the slight experimental drop in $M_s(T)$ below the Verwey transition obtained in **Figure S2** has been previously described in bulk magnetite [64]. By using the values obtained from our calculations for $K_{dip}$ and $K_{sh-uni}$ at 300 K as starting points, and considering that the thermal evolution of both anisotropies should be proportional to $M_S^2(T)$, we have applied the same Bloch-like law for the shape and dipolar contributions, thereby obtaining the following expressions:

$$K_{dip}(T) = K_{dip0}\left[1 - \left(\frac{T}{T_0}\right)^{\alpha_B}\right]^2 \quad (5)$$

$$K_{sh-uni}(T) = K_{sh-uni0}\left[1 - \left(\frac{T}{T_0}\right)^{\alpha_B}\right]^2 \quad (6)$$

where $K_{dip0}$ and $K_{sh-uni0}$ are the extrapolated dipolar and shape anisotropy constant values at 0 K, respectively.

This is more clearly illustrated in **Figure 6**, where the thermal evolution of the shape, dipolar, and magnetocrystalline to the effective uniaxial anisotropy constant, $K_{uni}$, have been represented separately for both species. As depicted, for the shape and dipolar contributions, the change between 300 K and 5 K is very small, < 10%. By subtracting both shape and dipolar contributions from $K_u$, we can obtain the uniaxial magnetocrystalline anisotropy contribution for each species. As depicted, the uniaxial magnetocrystalline anisotropy contributions for both MTB species follow a qualitatively similar evolution, much more complex than the shape and dipolar counterparts. Above $T_v$, the uniaxial magnetocrystalline contribution is null, as expected, but below $T_v$, it progressively increases with decreasing temperature following a non-monotonic trend. First it rapidly increases up to ~8-10 kJ m$^{-3}$ at 70 K, then it describes a shoulder, reaching ~12-14 kJ m$^{-3}$ at 40 K, and finally it increases again up to ~22-24 kJ m$^{-3}$ at 5 K.

This behavior clearly differs from the one obtained in stoichiometric bulk magnetite, in which the transition from a cubic to a monoclinic structure is sharp, and the monoclinic phase exhibits a uniaxial magnetocrystalline anisotropy constant of ~27 kJ m$^{-3}$ at 5 K [51,65]. In the case of magnetosomes, only at 5 K, the magnetocrystalline anisotropy contribution practically recovers the bulk magnetite value. Therefore, these results indicate that in magnetosomes and other similar high-quality magnetite-based nanoparticles, instead of having an abrupt Verwey transition in a very narrow range of temperature, the transition is slower and more complex, as described before. Interestingly, this evolution of the uniaxial magnetocrystalline anisotropy resembles the one from the ZFC curves (see **Figure S3**), suggesting that the "low T transition" appearing around 30-50 K in these and other magnetite-based nanoparticles could be linked to the change in the uniaxial magnetocrystalline anisotropy evolution observed around the same range of temperatures. The reasons behind this complex behavior could be related to the presence of a distribution of defects, vacancies, and/or doping elements in the crystalline structure of these nanoparticles. It is well-known that magnetite is greatly affected but tiny structural modifications [60,71]. These structural modifications could contribute to a progressive change in the crystalline structure, from cubic to monoclinic with decreasing temperature, although further experiments would be needed to shed more light on this matter.

It is important to emphasize that this distinction of individual contributions from experimental measurements represents a significant achievement, as such differentiation is often exceedingly difficult with other magnetite-based nanoparticles. This underscores the exceptional utility of magnetosomes as ideal model magnetic nanoparticles.

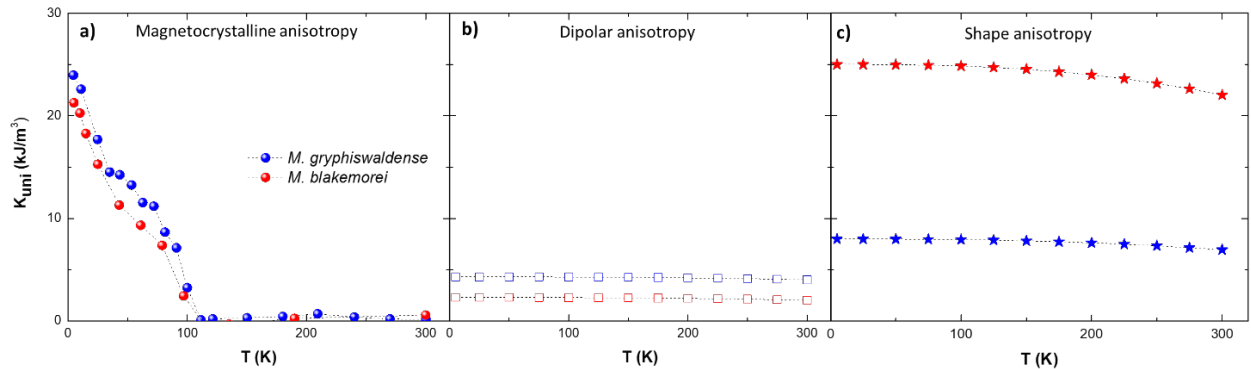

**Figure 6.** Thermal evolution of the a) magnetocrystalline, b) dipolar and c) shape anisotropies to the effective uniaxial anisotropy constant, $K_{uni}$, for *M. gryphiswaldense* and *M. blakemorei*.

## 4. Conclusion

In this study, we successfully isolated and analyzed the diverse anisotropic contributions affecting the magnetic properties of magnetite-based nanoparticles. Using magnetosomes from two MTB species, *M. gryphiswaldense* and *M. blakemorei*, which display similar composition, dimensions and crystallinity, but very different shape ($K_{sh}$ ~7 kJ/m³ and ~23 kJ/m³ at 300 K, respectively), we conducted a comparative analysis while maintaining consistency in other influential parameters. The dipolar interactions contributed a minor uniaxial anisotropy of approximately 2-4 kJ/m³ at 300 K. Despite minimal thermal variations in both shape and dipolar anisotropies across the examined temperature range, the differences in shape anisotropy between both species markedly influenced the *M* vs *H* loops, notably enhancing the coercivity for *M. blakemorei* by up to two-fold compared to *M. gryphiswaldense*. This underscores the critical role of shape anisotropy in modifying magnetic hysteresis in nanoparticles, which is especially relevant for applications of these nanoparticles in different fields, including biomedicine.

Furthermore, the uniaxial contribution of magnetocrystalline anisotropy was found to predominantly affect the magnetic behavior at lower temperatures. Through computations using a dynamic Stoner-Wohlfarth model, we were able to distinctly separate and trace the thermal evolution of the various anisotropic contributions: shape, dipolar, and magnetocrystalline. Notably, the intricate magnetic behavior observed in magnetosomes and other magnetite-based nanoparticles below the Verwey transition ($T_v$ ~110-120 K) is governed by a non-monotonic increase in uniaxial magnetocrystalline anisotropy as temperature decreases. This triggers a progressive Verwey transition, with the magnetite structure transitioning from cubic to monoclinic, nearly culminating at temperatures around 5 K, where the uniaxial magnetocrystalline anisotropy (~22-24 kJ/m³) approximates the values found in bulk magnetite. Despite the similar behavior in both species, our method has allowed us to pinpoint small differences in the magnetocrystalline anisotropy of their magnetosomes.

These findings highlight the exceptional utility of magnetosomes as ideal model nanoparticles for detailed studies, where a combination of experimental measurements and computational simulations facilitates a comprehensive understanding of the different factors influencing the magnetic responses of nanoparticles. This work not only enhances our understanding of magnetic nanoparticle behavior but also demonstrates the potential applications of magnetosomes in nanomagnetic research.

**CRediT authorship contribution statement**

- **David Gandia:** Investigation, Methodology, Formal Analysis, Visualization, Writing - Review & Editing.
- **Lourdes Marcano:** Investigation, Formal Analysis, Visualization, Writing - Review & Editing.
- **Lucía Gandarias:** Investigation, Methodology, Formal Analysis, Visualization, Writing - Review & Editing.
- **Alicia G. Gubieda:** Investigation, Writing - Review & Editing.
- **Ana García Prieto:** Conceptualization, Project Administration, Funding Acquisition, Writing - Review & Editing.
- **Luis Fernández Barquín:** Project Administration, Funding Acquisition, Resources, Supervision, Writing - Review & Editing.
- **José Ignacio Espeso:** Investigation, Data Curation.
- **Elizabeth Martín Jefremovas:** Investigation, Writing - Review & Editing.
- **Iñaki Orue:** Resources, Validation, Methodology, Investigation.

- **Ana Abad Díaz de Cerio:** Resources, Validation, Conceptualization, Supervision.
- **Mª Luisa Fdez-Gubieda:** Conceptualization, Project Administration, Funding Acquisition, Resources, Validation, Supervision, Writing - Review & Editing.
- **Javier Alonso:** Conceptualization, Project Administration, Funding Acquisition, Formal Analysis, Writing - Original Draft, Visualization, Writing - Review & Editing.

**Declaration of Generative AI and AI-assisted technologies in the writing process**

During the preparation of this work the author(s) used ChatGPT in order to correct errors in the use of English and improve grammar. After using this tool/service, the author(s) reviewed and edited the content as needed and take(s) full responsibility for the content of the publication.

**Declaration of Competing Interest**

The authors declare that they have no known competing financial interests or personal relationships that could have appeared to influence the work reported in this paper.

Acknowledgments

This work has been funded by the Spanish Government (grant PID2020-115704RB-C3 funded by MCIN/AEI/10.13039/501100011033) and the Basque Government (grant IT1479-22). L.G. would like to acknowledge the financial support provided through a postdoctoral fellowship from the Basque Government (POS_2022_1_0017). L.M. thanks the Horizon Europe Programme for the financial support provided through a Marie Sklodowska-Curie fellowship (101067742) and the BBVA Foundation for the Leonardo Fellowships for Researchers and Cultural Creators 2022. The authors thank SGIker (UPV/EHU/ERDF, EU) for technical and human support.

References

[1]  R. Blakemore, D.A. Bazylinski, C.T. Lefèvre, D. Schüler, Magnetotactic bacteria, in: Science (1979), Annual Reviews 4139 El Camino Way, P.O. Box 10139, Palo Alto, CA 94303-0139, USA, 2013: pp. 377–379. https://doi.org/10.1126/science.170679.

[2]  D.A. Bazylinski, T.J. Williams, C.T. Lefèvre, D. Trubitsyn, J. Fang, T.J. Beveridge, B.M. Moskowitz, B. Ward, S. Schübbe, B.L. Dubbels, B. Simpson, Magnetovibrio blakemorei gen. nov., sp. nov., a magnetotactic bacterium (Alphaproteobacteria: Rhodospirillaceae) isolated from a salt marsh, Int J Syst Evol Microbiol 63 (2013) 1824–1833. https://doi.org/10.1099/ijs.0.044453-0.

[3]  D.A. Bazylinski, R.B. Frankel, Magnetosome formation in prokaryotes, Nat Rev Microbiol 2 (2004) 217–230. https://doi.org/10.1038/nrmicro842.

[4]  A. Komeili, H. Vali, T.J. Beveridge, D.K. Newman, Magnetosome vesicles are present before magnetite formation, and MamA is required for their activation, Proc Natl Acad Sci U S A 101 (2004) 3839–3844. https://doi.org/10.1073/PNAS.0400391101.

[5]  R.B. Frankel, R.E. Dunin-Borkowski, M. Pósfai, D.A. Bazylinski, Magnetic Microstructure of Magnetotactic Bacteria, in: Handbook of Biomineralization: Biological Aspects and Structure Formation, 2008: pp. 126–144. https://doi.org/10.1002/9783527619443.ch8.

[6]  D. Gandia, L. Gandarias, I. Rodrigo, J. Robles-García, R. Das, E. Garaio, J.Á. García, M.H. Phan, H. Srikanth, I. Orue, J. Alonso, A. Muela, M.L. Fdez-Gubieda, Unlocking the Potential of Magnetotactic Bacteria as Magnetic Hyperthermia Agents, Small 15 (2019) 1902626. https://doi.org/10.1002/smll.201970222.

[7]  M.L. Fdez-Gubieda, J. Alonso, A. García-Prieto, A. García-Arribas, L. Fernández Barquín, A. Muela, Magnetotactic bacteria for cancer therapy, J Appl Phys 128 (2020) 070902. https://doi.org/10.1063/5.0018036.

[8]  E. Alphandéry, Applications of Magnetosomes Synthesized by Magnetotactic Bacteria in Medicine, Front Bioeng Biotechnol 2 (2014) 5. https://doi.org/10.3389/fbioe.2014.00005.

[9]  N. Mokrani, O. Felfoul, F. Afkhami Zarreh, M. Mohammadi, R. Aloyz, G. Batist, S. Martel, Magnetotactic bacteria penetration into multicellular tumor spheroids for targeted therapy., in: 2010 Annual International Conference of the IEEE Engineering in Medicine and Biology, 2010: pp. 4371–4374.

[10]   Z. Xiang, X. Yang, J. Xu, W. Lai, Z. Wang, Z. Hu, J. Tian, L. Geng, Q. Fang, Tumor detection using magnetosome nanoparticles functionalized with a newly screened EGFR/HER2 targeting peptide, Biomaterials 115 (2017) 53–64. https://doi.org/10.1016/j.biomaterials.2016.11.022.

[11]   S. Martel, M. Mohammadi, O. Felfoul, Zhao Lu, P. Pouponneau, Flagellated Magnetotactic Bacteria as Controlled MRI-trackable Propulsion and Steering Systems for Medical Nanorobots Operating in the Human Microvasculature, Int J Rob Res 28 (2009) 571–582. https://doi.org/10.1177/0278364908100924.

[12]   A.S. Mathuriya, Magnetotactic bacteria for cancer therapy, Biotechnol Lett 37 (2015) 491–498. https://doi.org/10.1007/s10529-014-1728-6.

[13]   G. Vargas, J. Cypriano, T. Correa, P. Leão, D. Bazylinski, F. Abreu, Applications of Magnetotactic Bacteria, Magnetosomes and Magnetosome Crystals in Biotechnology and Nanotechnology: Mini-Review, Molecules 23 (2018) 2438. https://doi.org/10.3390/molecules23102438.

[14]   A. Muela, D. Muñoz, R. Martín-Rodríguez, I. Orue, E. Garaio, A. Abad Díaz de Cerio, J. Alonso, J.Á. García, M.L. Fdez-Gubieda, Optimal Parameters for Hyperthermia Treatment Using Biomineralized Magnetite Nanoparticles: Theoretical and Experimental Approach, The Journal of Physical Chemistry C 120 (2016) 24437–24448. https://doi.org/10.1021/acs.jpcc.6b07321.

[15]   M. Marmol, E. Gachon, D. Faivre, Colloquium: Magnetotactic bacteria: From flagellar motor to collective effects, Rev Mod Phys 96 (2024) 021001. https://doi.org/10.1103/REVMODPHYS.96.021001.

[16]   J.F. Rupprecht, N. Waisbord, C. Ybert, C. Cottin-Bizonne, L. Bocquet, Velocity Condensation for Magnetotactic Bacteria, Phys Rev Lett 116 (2016) 168101. https://doi.org/10.1103/PHYSREVLETT.116.168101.

[17]   A. Petroff, A. Rosselli-Calderon, B. Roque, P. Kumar, Phases of active matter composed of multicellular magnetotactic bacteria near a hard surface, Phys Rev Fluids 7 (2022) 053102. https://doi.org/10.1103/PHYSREVFLUIDS.7.053102.

[18]   T. Gwisai, N. Mirkhani, M.G. Christiansen, T.T. Nguyen, V. Ling, S. Schuerle, Magnetic torque-driven living microrobots for increased tumor infiltration, Sci Robot 7 (2022). https://doi.org/10.1126/SCIROBOTICS.ABO0665.

[19]   N. Mirkhani, M.G. Christiansen, T. Gwisai, S. Menghini, S. Schuerle, Spatially selective delivery of living magnetic microrobots through torque-focusing, Nat Commun 15 (2024) 1–14. https://doi.org/10.1038/s41467-024-46407-4.

[20]   E.M. Jefremovas, L. Gandarias, I. Rodrigo, L. Marcano, C. Gruttner, J.A. Garcia, E. Garayo, I. Orue, A. Garcia-Prieto, A. Muela, M.L. Fernandez-Gubieda, J. Alonso, L.F. Barquin, Nanoflowers versus magnetosomes: Comparison between two promising candidates for magnetic hyperthermia therapy, IEEE Access 9 (2021) 99552–99561. https://doi.org/10.1109/ACCESS.2021.3096740.

[21]   L. Gandarias, E.M. Jefremovas, D. Gandia, L. Marcano, V. Martínez-Martínez, P. Ramos-Cabrer, D.M. Chevrier, S. Valencia, L. Fernández Barquín, M.L. Fdez-Gubieda, J. Alonso, A. García-Prieto,

A. Muela, Incorporation of Tb and Gd improves the diagnostic functionality of magnetotactic bacteria, Mater Today Bio 20 (2023) 100680. https://doi.org/10.1016/J.MTBIO.2023.100680.

[22] I. Orue, L. Marcano, P. Bender, A. García-Prieto, S. Valencia, M.A. Mawass, D. Gil-Cartón, D. Alba Venero, D. Honecker, A. García-Arribas, L. Fernández Barquín, A. Muela, M.L. Fdez-Gubieda, Configuration of the magnetosome chain: A natural magnetic nanoarchitecture, Nanoscale 10 (2018) 7407–7419. https://doi.org/10.1039/c7nr08493e.

[23] L. Marcano, I. Orue, D. Gandia, L. Gandarias, M. Weigand, R.M. Abrudan, A. García-Prieto, A. García-Arribas, A. Muela, M.L. Fdez-Gubieda, S. Valencia, Magnetic Anisotropy of Individual Nanomagnets Embedded in Biological Systems Determined by Axi-asymmetric X-ray Transmission Microscopy, ACS Nano 16 (2022) 7398–7408. https://doi.org/10.1021/ACSNANO.1C09559.

[24] E.M. Jefremovas, L. Gandarias, L. Marcano, A. Gacía-Prieto, I. Orue, A. Muela, M.L. Fdez-Gubieda, L.F. Barquín, J. Alonso, Modifying the magnetic response of magnetotactic bacteria: incorporation of Gd and Tb ions into the magnetosome structure, Nanoscale Adv 4 (2022) 2649–2659. https://doi.org/10.1039/d2na00094f.

[25] D. Gandia, L. Gandarias, L. Marcano, I. Orue, D. Gil-Cartón, J. Alonso, A. García-Arribas, A. Muela, M.L. Fdez-Gubieda, Elucidating the role of shape anisotropy in faceted magnetic nanoparticles using biogenic magnetosomes as a model, Nanoscale 12 (2020) 16081–16090. https://doi.org/10.1039/d0nr02189j.

[26] D. Gandia, L. Gandarias, I. Rodrigo, J. Robles-García, R. Das, E. Garaio, J.Á. García, M.H. Phan, H. Srikanth, I. Orue, J. Alonso, A. Muela, M.L. Fdez-Gubieda, Unlocking the Potential of Magnetotactic Bacteria as Magnetic Hyperthermia Agents, Small 15 (2019) 1902626. https://doi.org/10.1002/smll.201970222.

[27] Z. Nemati, R. Das, J. Alonso, E. Clements, M.H. Phan, H. Srikanth, Iron Oxide Nanospheres and Nanocubes for Magnetic Hyperthermia Therapy: A Comparative Study, J Electron Mater 46 (2017) 3764–3769. https://doi.org/10.1007/s11664-017-5347-6.

[28] R. Das, J. Alonso, Z. Nemati Porshokouh, V. Kalappattil, D. Torres, M.-H. Phan, E. Garaio, J.Á. García, J.L. Sanchez Llamazares, H. Srikanth, Tunable High Aspect Ratio Iron Oxide Nanorods for Enhanced Hyperthermia, The Journal of Physical Chemistry C 120 (2016) 10086–10093. https://doi.org/10.1021/acs.jpcc.6b02006.

[29] Z. Nemati, S.M. Salili, J. Alonso, A. Ataie, R. Das, M.H. Phan, H. Srikanth, Superparamagnetic iron oxide nanodiscs for hyperthermia therapy: Does size matter?, J Alloys Compd 714 (2017) 709–714. https://doi.org/https://doi.org/10.1016/j.jallcom.2017.04.211.

[30] Z. Nemati, J. Alonso, I. Rodrigo, R. Das, E. Garaio, J.Á. García, I. Orue, M.H. Phan, H. Srikanth, Improving the Heating Efficiency of Iron Oxide Nanoparticles by Tuning Their Shape and Size, Journal of Physical Chemistry C 122 (2018) 2367–2381. https://doi.org/10.1021/acs.jpcc.7b10528.

[31] R.M. Fratila, S. Rivera-Fernández, J.M. De La Fuente, Shape matters: Synthesis and biomedical applications of high aspect ratio magnetic nanomaterials, Nanoscale 7 (2015) 8233–8260. https://doi.org/10.1039/c5nr01100k.

[32] C.T. Lefèvre, D.A. Bazylinski, C.T. Lefevre, D.A. Bazylinski, Ecology, Diversity, and Evolution of Magnetotactic Bacteria, Microbiology and Molecular Biology Reviews 77 (2013) 497–526. https://doi.org/10.1128/MMBR.00021-13.

[33] A. de Souza Cabral, M. Verdan, R. Presciliano, F. Silveira, T. Correa, F. Abreu, Large-Scale Cultivation of Magnetotactic Bacteria and the Optimism for Sustainable and Cheap Approaches in Nanotechnology, Mar Drugs 21 (2023) 60. https://doi.org/10.3390/MD21020060.

[34] R. Amann, J. Peplies, D. Schüler, Diversity and Taxonomy of Magnetotactic Bacteria, in: Magnetoreception and Magnetosomes in Bacteria, Springer, Berlin, Heidelberg , 2006: pp. 25–36. https://doi.org/10.1007/7171_037.

[35] L. Le Nagard, V. Morillo-López, C. Fradin, D.A. Bazylinski, Growing Magnetotactic Bacteria of the Genus Magnetospirillum: Strains MSR-1, AMB-1 and MS-1, Journal of Visualized Experiments 140 (2018) e58536. https://doi.org/10.3791/58536.

[36] T. Matsunaga, T. Sakaguchi, F. Tadakoro, Magnetite formation by a magnetic bacterium capable of growing aerobically, Appl Microbiol Biotechnol 35 (1991) 651–655. https://doi.org/10.1007/BF00169632.

[37] K.H. Schleifer, D. Schüler, S. Spring, M. Weizenegger, R. Amann, W. Ludwig, M. Köhler, The Genus Magnetospirillum gen. nov. Description of Magnetospirillum gryphiswaldense sp. nov. and Transfer of Aquaspirillum magnetotacticum to Magnetospirillum magnetotacticum comb. nov., Syst Appl Microbiol 14 (1991) 379–385. https://doi.org/10.1016/S0723-2020(11)80313-9.

[38] D.A. Bazylinski, R.B. Frankel, H.W. Jannasch, Anaerobic magnetite production by a marine, magnetotactic bacterium, Nature 334 (1988) 518–519. https://doi.org/10.1038/334518a0.

[39] A.C. V. Araujo, F. Abreu, K.T. Silva, D.A. Bazylinski, U. Lins, Magnetotactic bacteria as potential sources of bioproducts, Mar Drugs 13 (2015) 389. https://doi.org/10.3390/md13010389.

[40] K.T. Silva, P.E. Leão, F. Abreu, J.A. López, M.L. Gutarra, M. Farina, D.A. Bazylinski, D.M.G. Freire, U. Lins, Optimization of Magnetosome Production and Growth by the Magnetotactic Vibrio Magnetovibrio blakemorei Strain MV-1 through a Statistics-Based Experimental Design, Appl Environ Microbiol 79 (2013) 2823. https://doi.org/10.1128/AEM.03740-12.

[41] S.S. Kalirai, D.A. Bazylinski, A.P. Hitchcock, Anomalous Magnetic Orientations of Magnetosome Chains in a Magnetotactic Bacterium: Magnetovibrio blakemorei Strain MV-1, PLoS One 8 (2013) e53368. https://doi.org/10.1371/JOURNAL.PONE.0053368.

[42] L. Jovane, F. Florindo, D.A. Bazylinski, U. Lins, Prismatic magnetite magnetosomes from cultivated Magnetovibrio blakemorei strain MV-1: A magnetic fingerprint in marine sediments?, Environ Microbiol Rep 4 (2012) 664–668. https://doi.org/10.1111/1758-2229.12000.

[43] D.A. Bazylinski, T.J. Williams, C.T. Lefèvre, D. Trubitsyn, J. Fang, T.J. Beveridge, B.M. Moskowitz, B. Ward, S. Schübbe, B.L. Dubbels, B. Simpson, Magnetovibrio blakemorei gen. nov., sp. nov., a

magnetotactic bacterium (Alphaproteobacteria: Rhodospirillaceae) isolated from a salt marsh, Int J Syst Evol Microbiol 63 (2013) 1824–1833. https://doi.org/10.1099/IJS.0.044453-0.

[44] B.W. Zingsem, T. Feggeler, A. Terwey, S. Ghaisari, D. Spoddig, D. Faivre, R. Meckenstock, M. Farle, M. Winklhofer, Biologically encoded magnonics, Nat Commun 10 (2019) 1–8. https://doi.org/10.1038/s41467-019-12219-0.

[45] U. Heyen, D. Schüler, Growth and magnetosome formation by microaerophilic Magnetospirillum strains in an oxygen-controlled fermentor, Appl Microbiol Biotechnol 61 (2003) 536–544. https://doi.org/10.1007/s00253-002-1219-x.

[46] C.A. Schneider, W.S. Rasband, K.W. Eliceiri, NIH Image to ImageJ: 25 years of image analysis, Nat Methods 9 (2012) 671–675. https://doi.org/10.1038/nmeth.2089.

[47] F. Walz, The Verwey transition - a topical review, Journal of Physics: Condensed Matter 14 (2002) R285–R340. https://doi.org/10.1088/0953-8984/14/12/203.

[48] J. García, G. Subías, The Verwey transition - A new perspective, Journal of Physics Condensed Matter 16 (2004) R145–R178. https://doi.org/10.1088/0953-8984/16/7/R01.

[49] E.J.W. Verwey, Electronic Conduction of Magnetite (Fe3O4) and its Transition Point at Low Temperatures, Nature 144 (1939) 327–328. https://doi.org/10.1038/144327b0.

[50] I. Castellanos-Rubio, O. Arriortua, D. Iglesias-Rojas, A. Barón, I. Rodrigo, L. Marcano, J.S. Garitaonandia, I. Orue, M.L. Fdez-Gubieda, M. Insausti, A Milestone in the Chemical Synthesis of Fe3O4 Nanoparticles: Unreported Bulklike Properties Lead to a Remarkable Magnetic Hyperthermia, Chemistry of Materials 33 (2021) 8693–8704. https://doi.org/10.1021/acs.chemmater.1c02654.

[51] R. Řezníček, V. Chlan, H. Štěpánková, P. Novák, M. Maryško, Magnetocrystalline anisotropy of magnetite, Journal of Physics Condensed Matter 24 (2012) 055501. https://doi.org/10.1088/0953-8984/24/5/055501.

[52] L. Marcano, A. García-Prieto, D. Muñoz, L. Fernández Barquín, I. Orue, J. Alonso, A. Muela, M.L. Fdez-Gubieda, Influence of the bacterial growth phase on the magnetic properties of magnetosomes synthesized by Magnetospirillum gryphiswaldense, Biochim Biophys Acta Gen Subj 1861 (2017) 1507–1514. https://doi.org/10.1016/j.bbagen.2017.01.012.

[53] E.M. Jefremovas, L. Gandarias, L. Marcano, A. Gacía-Prieto, I. Orue, A. Muela, M.L. Fdez-Gubieda, L.F. Barquín, J. Alonso, Modifying the magnetic response of magnetotactic bacteria: incorporation of Gd and Tb ions into the magnetosome structure, Nanoscale Adv 4 (2022) 2649–2659. https://doi.org/10.1039/d2na00094f.

[54] T. Kim, S. Sim, S. Lim, M.A. Patino, J. Hong, J. Lee, T. Hyeon, Y. Shimakawa, S. Lee, J.P. Attfield, J.G. Park, Slow oxidation of magnetite nanoparticles elucidates the limits of the Verwey transition, Nat Commun 12 (2021) 1–12. https://doi.org/10.1038/s41467-021-26566-4.

[55] R. Das, V. Kalappattil, M.H. Phan, H. Srikanth, Magnetic anomalies associated with domain wall freezing and coupled electron hopping in magnetite nanorods, J Magn Magn Mater 522 (2021) 167564. https://doi.org/10.1016/j.jmmm.2020.167564.

[56]  S. Sahling, J.E. Lorenzo, G. Remenyi, C. Marin, V.L. Katkov, V.A. Osipov, Heat capacity signature of frustrated trimerons in magnetite, Sci Rep 10 (2020) 1–6. https://doi.org/10.1038/s41598-020-67955-x.

[57]  M.S. Senn, J.P. Wright, J.P. Attfield, Charge order and three-site distortions in the Verwey structure of magnetite, Nature 481 (2012) 173–176. https://doi.org/10.1038/nature10704.

[58]  V. Brabers, F. Walz, H. Kronmüller, Impurity effects upon the Verwey transition in magnetite, Phys Rev B Condens Matter Mater Phys 58 (1998) 14163–14166. https://doi.org/10.1103/PhysRevB.58.14163.

[59]  V. Skumryev, H.J. Blythe, J. Cullen, J.M.D. Coey, AC susceptibility of a magnetite crystal, J Magn Magn Mater 196 (1999) 515–517. https://doi.org/10.1016/S0304-8853(98)00863-4.

[60]  H. Kronmüller, F. Walz, Magnetic after-effects in Fe3O4 and vacancy-doped magnetite, Philosophical Magazine B 42 (1980) 433–452. https://doi.org/10.1080/01418638008221886.

[61]  B.D. Cullity, C.D. Graham, Introduction to Magnetic Materials (2nd Edition), 2009. https://doi.org/10.1016/S1369-7021(09)70091-4.

[62]  Z. Boekelheide, J.T. Miller, C. Grüttner, C.L. Dennis, The effects of intraparticle structure and interparticle interactions on the magnetic hysteresis loop of magnetic nanoparticles, J Appl Phys 126 (2019) 043903. https://doi.org/10.1063/1.5094180.

[63]  Z. Ma, J. Mohapatra, K. Wei, J.P. Liu, S. Sun, Magnetic Nanoparticles: Synthesis, Anisotropy, and Applications, Chem Rev 123 (2023) 3904–3943. https://doi.org/10.1021/acs.chemrev.1c00860.

[64]  Ö. Özden, Coercive force of single crystals of magnetite at low temperatures, Geophys J Int 141 (2000) 351–356. http://dx.doi.org/10.1046/j.1365-246x.2000.00081.x.

[65]  K. Abe, Y. Miyamoto, S. Chikazumi, Magnetocrystalline Anisotropy of Low Temperature Phase of Magnetite, J Physical Soc Japan 41 (1976) 1894–1902. https://doi.org/10.1143/JPSJ.41.1894.

[66]  G. Barrera, P. Tiberto, C. Sciancalepore, M. Messori, F. Bondioli, P. Allia, Verwey transition temperature distribution in magnetic nanocomposites containing polydisperse magnetite nanoparticles, J Mater Sci 54 (2019) 8346–8360. https://doi.org/10.1007/s10853-019-03510-y.

[67]  G.F. Goya, T.S. Berquó, F.C. Fonseca, M.P. Morales, Static and dynamic magnetic properties of spherical magnetite nanoparticles, J Appl Phys 94 (2003) 3520–3528. https://doi.org/10.1063/1.1599959.

[68]  E.C. Stoner, E.P. Wohlfarth, A mechanism of magnetic hysteresis in heterogeneous alloys, Philosophical Transactions of the Royal Society A: Mathematical, Physical and Engineering Sciences 240 (1948) 599–642. https://doi.org/10.1109/TMAG.1991.1183750.

[69]  L. Marcano, D. Muñoz, R. Martín-Rodríguez, I. Orue, J. Alonso, A. García-Prieto, A. Serrano, S. Valencia, R. Abrudan, L. Fernández Barquín, A. García-Arribas, A. Muela, M.L.L. Fdez-Gubieda, Magnetic Study of Co-Doped Magnetosome Chains, Journal of Physical Chemistry C 122 (2018) 7541–7550. https://doi.org/10.1021/acs.jpcc.8b01187.

[70] J. Carrey, B. Mehdaoui, M. Respaud, Simple models for dynamic hysteresis loop calculations of magnetic single-domain nanoparticles: Application to magnetic hyperthermia optimization, J Appl Phys 109 (2011) 083921. https://doi.org/10.1063/1.3551582.

[71] D. González-Alonso, J. González, H. Gavilán, J. Fock, L. Zeng, K. Witte, P. Bender, L.F. Barquín, C. Johansson, Revealing a masked Verwey transition in nanoparticles of coexisting Fe-oxide phases, RSC Adv 11 (2020) 390–396. https://doi.org/10.1039/d0ra09226f.